# Experimental momentum spectra of identified hadrons in jets and the predictions from LPHD + MLLA

N.C. Brümmer
NIKHEF-H
P.O. Box 41882
1009 DB Amsterdam NL
e-mail: Nichol.Brummer@nikhef.nl

**Abstract**

Experimental data on the shape of hadronic momentum spectra are compared with theoretical predictions in the context of calculations in the Modified Leading Log Approximation (MLLA), under the assumption of Local Parton Hadron Duality (LPHD). Considered are experimental measurements at $\mathrm{e}^{+}\mathrm{e}^{-}$-colliders of $\xi_p^*$, the position of the maximum in the distribution of $\xi_p = \log(1/x_p)$, where $x_p = p/p_{\text{beam}}$. The parameter $\xi_p^*$ is determined for various hadrons at various centre of mass energies. It is interesting to look at the dependence of $\xi_p^*$ on the hadron type. This is used to study the influence of the hadron type on the cut-off scale $Q_0$ in the parton shower development. The dependence of $\xi_p^*$ on the centre of mass energy is seen to be described adequately by perturbation theory. The approach is made quantitative by extracting a value of $\alpha_s(\mathrm{m_Z})$ from an overall fit to the scaling behaviour of $\xi_p^*$.

# 1 Introduction

During a few years of LEP running, a large amount of information was collected on identified hadron species in jets. The higher centre of mass energy of LEP, compared to past $e^+e^-$ accelerators, also makes it easier to separate the behaviour of hadrons with a high momentum, which are correlated strongly to the primary quark, from those with a low momentum, created mainly during fragmentation and hadronisation.

Clear scaling violations have been observed in the shape of the charged particle $x_p$-distribution as function of the centre of mass energy. The strong coupling constant has been extracted from these scaling violations, using the behaviour at high momenta: $0.2 \lesssim x_p \lesssim 0.7$. Due to the larger statistical errors this is not possible for individually identified hadron species.

However, low momentum data for specific types of hadrons maybe be used to study the properties of jet-evolution and hadron-formation in the context of the LPHD hypothesis and MLLA calculations of parton spectra [1–4]. The assumption of 'Local Parton Hadron Duality' (LPHD) states that a calculated spectrum for 'partons' in a 'parton shower' can be related to the spectrum of real hadrons by simple normalisation constants. These constants have to be determined by experiment. A second assumption is that the low momentum part of the spectrum is not influenced in a significant way by hadrons that are correlated to the primary quark.

Calculations of the parton spectra in the 'Modified Leading Log Approximation' (MLLA) take into account next-to-leading logarithms in a consistent fashion. The physical mechanism relevant to these next-to-leading logarithms is the coherent emission of soft gluons inside a jet, leading to an angular ordering and an effective transverse momentum cutoff for the partons. Parton jets develop through repeated parton splittings, resulting in an increase of the multiplicity at lower momenta. The interplay of coherent emission of gluons and the creation of hadrons causes this spectrum to be cut off at very low momenta. Calculations predict the shape of the $\xi_p$ distribution. The resulting 'hump-backed' distribution is nearly gaussian. As an example, the Monte Carlo spectrum of the $\Lambda$ baryon in $Z^0$ decays can be seen in figures 1 and 2 . In the following $\xi_p^*$ will be determined for various types of hadrons and at various centre of mass energies. Subsequently a comparison is made with the theoretical calculations.

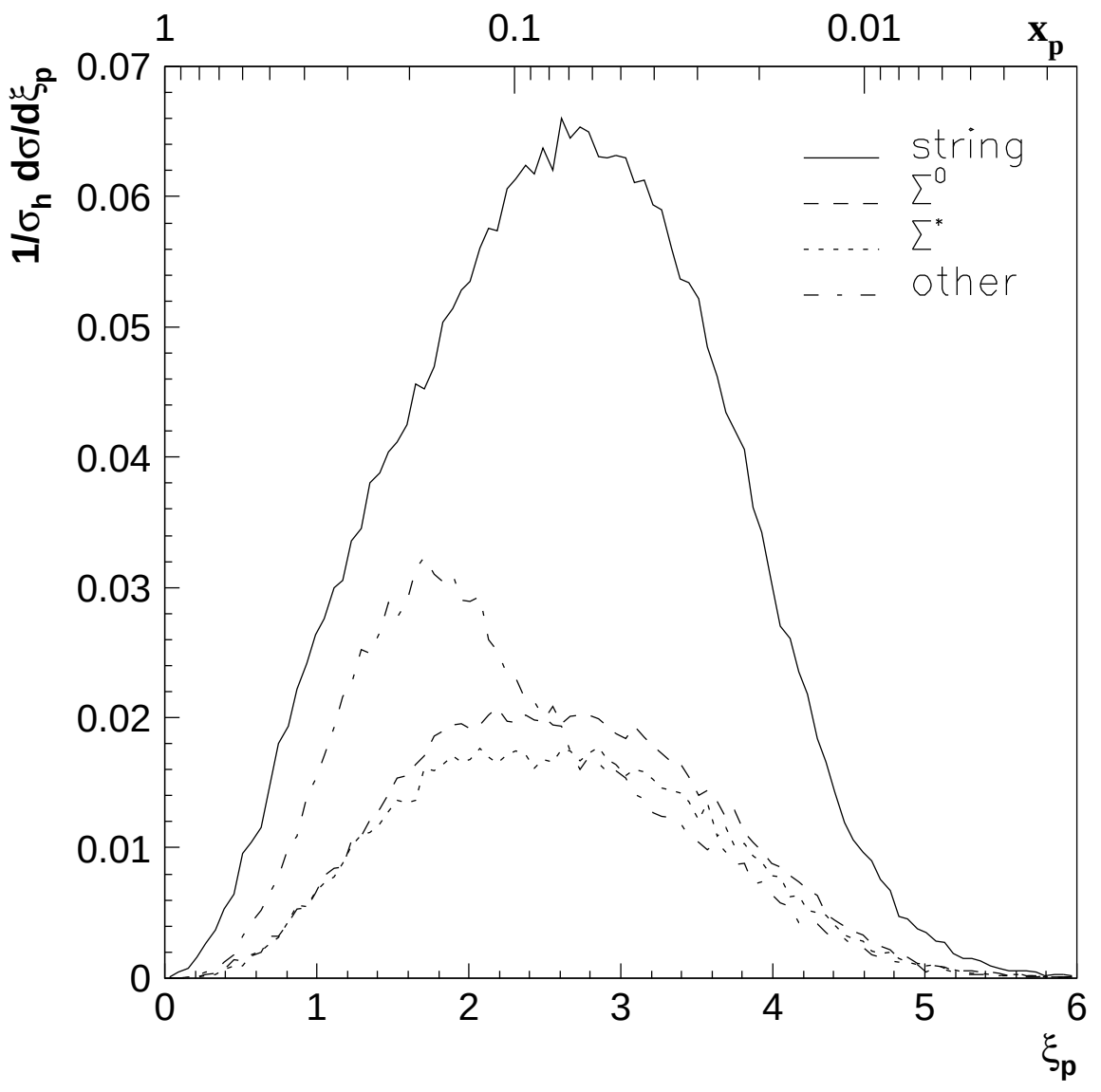

**Figure 1:** *The dependence of the $\Lambda$ spectrum on the species of its 'parent' particle in* JETSET*. Particles produced directly in the fragmentation process have the Lund-model string as 'parent'. The scales of both $x_p$ and $\xi_p$ are given for comparison.*

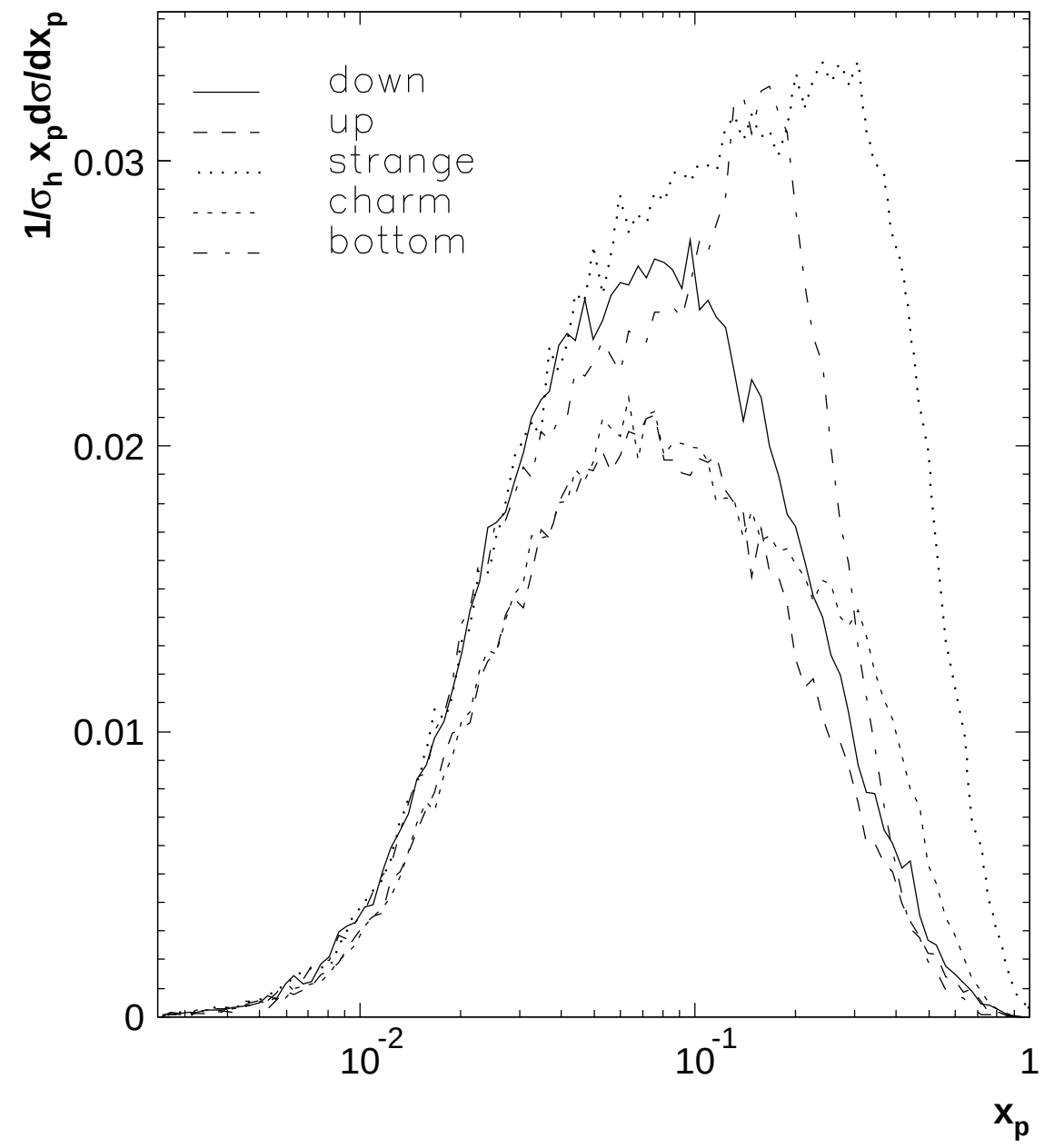

**Figure 2:** *The dependence of the $\Lambda$ spectrum on the flavour of the primary quarks in $Z^0$ decays at LEP, as predicted by* JETSET*. Notice that the high and low momentum regions are not separated for $\Lambda$'s in* s, c *and* b *events.*

| | $\sqrt{s}$ | charged | $\pi^\pm$ | $\pi^0$ | $K^\pm$ | $K^0$ | p | $\Lambda$ | $\Xi^-$ |
|---|---|---|---|---|---|---|---|---|---|
| ARGUS | $9.98$ | | $2.32 \pm 0.022$ | $2.42 \pm 0.08$ | $1.64 \pm 0.05$ | $1.72 \pm 0.04$ | $1.63 \pm 0.07$ | $1.44 \pm 0.034$ | $1.32 \pm 0.11$ |
| CLEO | $10.49$ | | $2.30 \pm 0.08$ | | | $1.68 \pm 0.17$ | $1.67 \pm 0.10$ | $1.47 \pm 0.08$ | |
| TASSO | $14$ | $2.45 \pm 0.05$ | $2.66 \pm 0.06$ | | $1.91 \pm 0.13$ | $1.65 \pm 0.13$ | $1.80 \pm 0.15$ | | |
| JADE | $14$ | | | $2.68 \pm 0.27$ | | | | | |
| TASSO | $22$ | $2.74 \pm 0.06$ | $2.99 \pm 0.09$ | | $2.41 \pm 0.23$ | $2.47 \pm 0.8$ | $2.14 \pm 0.27$ | $1.75 \pm 0.50$ | |
| JADE | $22.5$ | | | $2.84 \pm 0.30$ | | | | | |
| HRS | $29$ | | $> 3.3$ | | $2.28 \pm 0.42$ | $2.25 \pm 0.30$ | | $1.96 \pm 0.12$ | |
| TPC/$2\gamma$ | $29$ | | $3.00 \pm 0.05$ | | $2.14 \pm 0.05$ | $1.98 \pm 0.09$ | $2.13 \pm 0.11$ | | |
| TASSO | $34$ | $3.00 \pm 0.05$ | $3.17 \pm 0.05$ | | $2.35 \pm 0.20$ | $2.32 \pm 0.10$ | $2.24 \pm 0.09$ | $2.15 \pm 0.15$ | |
| JADE | $35$ | | | $3.24 \pm 0.05$ | | | | | |
| TASSO | $44$ | | $3.37 \pm 0.07$ | | $<3.5$ | | $<3.5$ | | |
| JADE | $44$ | | | $3.52 \pm 0.08$ | | | | | |
| TOPAZ | $58$ | $3.42 \pm 0.04$ | $3.51 \pm 0.07$ | | $2.83 \pm 0.10$ | | $2.64 \pm 0.06$ | | |
| OPAL | $91.2$ | $3.603 \pm 0.013$ | | | | $2.91 \pm 0.04$ | | $2.77 \pm 0.05$ | $2.57 \pm 0.11$ |
| DELPHI | $91.2$ | | | | | $2.62 \pm 0.11$ | | $2.81 \pm 0.04$ | |
| L3 | $91.2$ | $3.71 \pm 0.05$ | | $4.11 \pm 0.18$ | | | | | |

**Table 1:** *The values of $\xi_p^*$ for various hadrons as determined from the momentum spectra as measured by $e^+e^-$ experiments at different values of $\sqrt{s}$. For the $\eta$ meson, the L3 experiment [10] reported a value $\xi_p^* = 2.60 \pm 0.15$.*

Figure 1 shows that LPHD is not an obvious assumption: in the JETSET Monte Carlo, the spectrum of e.g. a $\Lambda$ baryon depends on its 'parent' particle. The LPHD assumption states that the sum of these spectra is proportional to the spectrum as calculated for a parton shower with a correctly chosen energy cut-off.

Moreover, the JETSET momentum spectrum of the $\Lambda$ baryon (figure 2) shows a strong dependence on the flavour of the primary quarks. For the heavier quarks, s, c and b, the high momentum contribution correlated to the primary quarks is not separated from the low momentum contribution related to the fragmentation process. This can be expected to cause shifts in $\xi_p^*$ that depend on the jet-flavour.

The spectrum of all charged particles at LEP has been fitted to the MLLA distribution, with free normalisation factors for pions, kaons and protons, depending on the centre of mass energy [5,6]. It was quite surprising [3] that the MLLA functions can also fit the high momentum part in the data for pions. This is probably a coincidence, since the heavier $K^0$ meson has a spectrum that can not be fitted as nicely at large $x_p$.

One way to see whether the observed shape of the spectrum is really due to coherent gluon emission is to compare experimental data to predictions of the JETSET Monte Carlo program, with the coherence (angular ordering) either turned on or turned off. The large dependence on the primary quark flavours and the fact that coherence in JETSET is not really necessary to fit the experimental data has led some to the conclusion [7] that coherence can not be demonstrated on the basis of hadron momentum spectra. Here an attempt will be made to infer more information using the properties of identified hadrons, concentrating on the observable $\xi_p^*$. This is done for data from LEP and from various other $e^+e^-$ colliders. Subsequently, the available results for $\xi_p^*$ are compared to the LPHD + MLLA approach [1,2].

## 2 Experimental values of $\xi_p^*$ for identified hadrons

At LEP, analyses of explicitly identified hadrons have been performed for the $\pi^0$ [6], the $K_S^0$ [8,9], the $\eta$ [10] and the $\Lambda$ [8,11,12,22]. At lower statistics data is also available for the $\Xi^-$ and $\Omega^-$ baryons, as well as the $\Sigma^*$ and $\Xi^*$ [8,11].

The values of $\xi_p^*$ correspond to low momenta, where the dependence on the primary quark flavours is expected to be small. To first approximation the distribution in $\xi_p$ is gaussian, but a distorted gaussian fits the calculated spectrum more accurately [13,2]. The present statistics of identified hadrons is not yet sufficient to extract the width and the two distortion parameters. However, since these other parameters depend more strongly on the high momentum data, they can be expected to depend stronger on the event flavour, which could bias the results. This is a reason not to use a distorted gaussian to determine $\xi_p^*$.

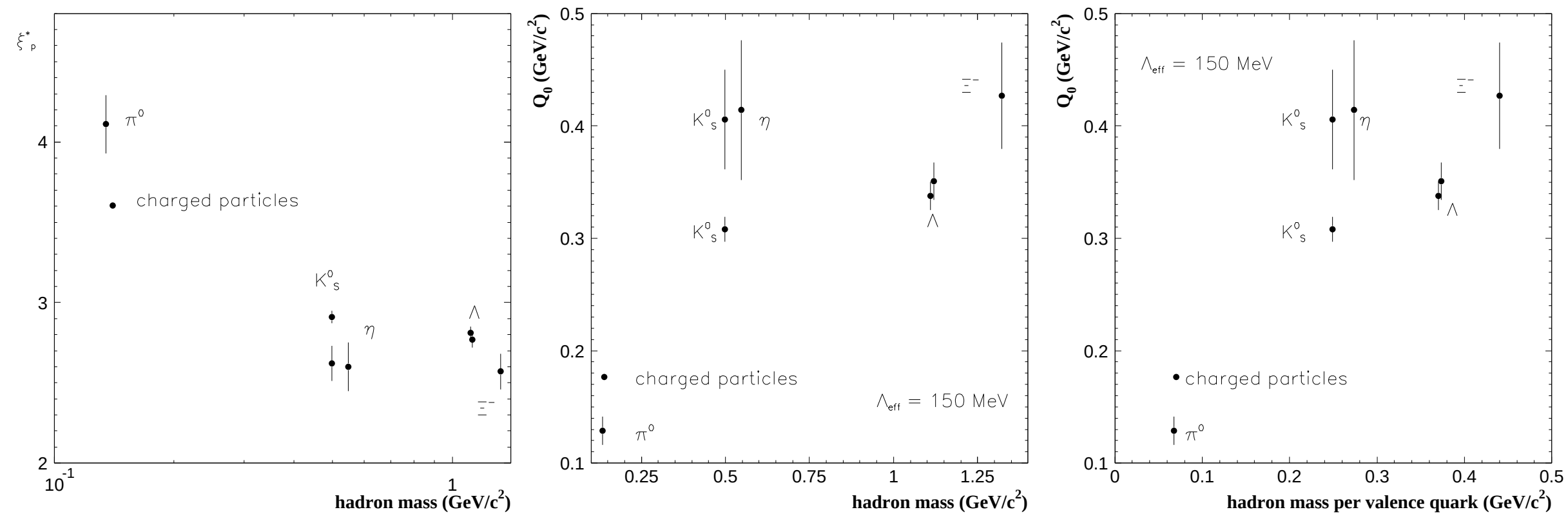

(a) $\xi_p^*$ versus the hadron mass (b) $Q_0$ versus the hadron mass (c) $Q_0$ versus the hadron mass per quark

**Figure 3:** *The maxima $\xi_p^*$ of the distributions in $\xi = -\log x_p$ for different hadrons as measured at LEP (table 1). the value of $\xi_p^*$ is a nearly linear function of $\log Q_0$. The exact dependence for $\Lambda_{eff} = 150$ MeV leads to figure (b), where the naive expectation would be that $Q_0$ is a linear function of the hadron mass. Figure (c) gives the dependence of $Q_0$ on the hadron mass divided by the number of valence quarks.*

In the following $\xi_p^*$ was determined by fitting the $\xi_p$ distribution in a limited range of $\xi_p$ around the maximum, to a gaussian distribution. A systematic error was estimated by changing the fitted range. The statistical and systematic errors were added in quadrature. The systematic error often dominated.

The results are given in table 1. Use was made of spectra published by CLEO [15], ARGUS [16], TPC [17], HRS [18], TASSO [19], JADE [20], TOPAZ [21], DELPHI [8,22], OPAL [5,9,11] and L3 [6,10]. Most of the recent publications reported values of $\xi_p^*$. Most experiments at lower centre of mass energies did not report the distribution of $\xi_p$ or $x_p$, but used the energy to define $x_E = 2E/E_{\text{beam}}$, or they divided by $\beta = v/c$ to obtain the 'scaling cross section'. Many different normalisations have been used, but fortunately this is unimportant for a determination $\xi_p^*$. The experiments (ARGUS and CLEO) near the $\Upsilon$ resonances give results for the 'continuum', and for the resonances itself, where the contributions from the continuum are subtracted. The values in table 1 refer to data from the continuum above the $\Upsilon$ resonance. In the data from ARGUS for charged pions and protons, the contributions from decays of $\mathrm{K^0_S}$ and $\Lambda$ were subtracted.

## 3 Predictions about $\xi_p^*$

The LPHD + MLLA calculations [2] of the distribution of $\xi_p$ depend on an effective QCD scale $\Lambda_{\text{eff}} \sim \Lambda_{\text{QCD}}$ and on a transverse momentum cutoff $Q_0$ in the evolution of the parton cascade. The value of $\xi_p^*$ is calculated to be a nearly linear function of $\log Q_0$.

The calculated dependence on the centre of mass energy is as follows [2]:

$$\xi_p^* = Y\left[\frac{1}{2} + \sqrt{C/Y} - C/Y + \mathcal{O}(Y^{-3/2})\right] + F(\lambda) \quad (1)$$

where $F(0) = 0$ and[1]

$$Y = \log(E_{\text{beam}}/\Lambda_{\text{eff}}), \qquad \lambda = \log(Q_0/\Lambda_{\text{eff}}) \quad (2)$$

and $\Lambda_{\text{eff}}$ is the 'effective' QCD scale, while only the cutoff scale $Q_0$ depends on the hadron type. The constant $C$ is calculated to be

$$C = \left(\frac{a}{4N_c}\right)^2 \frac{N_c}{b}, \quad (3)$$

[1]This detail [23] is not completely clear in reference [2], where equation (1) is given as a function of $Y - \lambda$, but in the approximation that $\lambda = 0$.

where

$$b = 11N_c/3 - 2n_f/3, \quad (4)$$
$$a = 11N_c/3 + 2n_f/(3N_c^2). \quad (5)$$

$N_c = 3$ is the number of colours and $n_f = 3$ is the active number of quark flavours in the fragmentation process. For $n_f = 3$ one finds $C = 0.2915$.

It is important to note that $F$ does not depend on $Y$, and that the first part of equation (1) is independent of $Q_0$. This predicted behaviour can be checked by comparing spectra of different identified particles. In the available momentum range this leads to a nearly linear dependence of $\xi_p^*$ on $Y$.

In ref. [2] various graphs show the results of numerical calculations for the dependence of various parameters on $E_{\text{beam}}$ and $Q_0 \neq \Lambda_{\text{eff}}$. Figure 4 of ref. [2] shows the dependence of $\xi_p^*$ on $Q_0$ for $\Lambda_{\text{eff}} = 150$ M$e$V, and at a number of beam energies. From this it is possible to extract $F$ after subtracting the $Y$ dependent part of equation (1). For an interpretation of the experimental values of $\xi_p^*$ it was useful to fit both this function $F(\lambda)$, and its inverse $\lambda(F)$ to polynomials. The result for $F(\lambda)$ is:

$$F(\lambda) = -1.380\cdot\lambda + 0.086\cdot\lambda^2 + 0.044\cdot\lambda^3 \pm 0.06 \quad (6)$$

while the inverse of $F$ is described by:

$$\lambda(F) = -0.749\cdot F - 0.054\cdot F^2 - 0.072\cdot F^3 \pm 0.06 \quad (7)$$

The given accuracy is the maximum deviation from the distribution plotted in ref. [2], in the range $-2 \lesssim F \lesssim 0$ and $0 \lesssim \lambda \lesssim 2$.

## 4 Dependence of $\xi_p^*$ on the hadron mass

The dependence of $\xi_p^*$ on the hadron mass as determined from LEP data is shown in figure 3(a). Using equation (7) we can convert the values of $\xi_p^*$ for a given value of $\Lambda_{\text{eff}} = 150$ M$e$V to values of $Q_0$. This result of this is shown in figure 3(b)

Naively one could expect that the cutoff scale $Q_0$ is larger for more massive hadrons. The most naive guess would be a linear dependence of the form $Q_0 \approx \Lambda_{\text{eff}} + m_{\text{hadron}}$. Figure 3(b) shows that this is not correct and that at least a separate treatment of mesons and baryons is necessary. A linear dependence of $Q_0$ on e.g. the hadron mass per constituent quark already describes the data slightly better, as can be seen in figure 3(c).

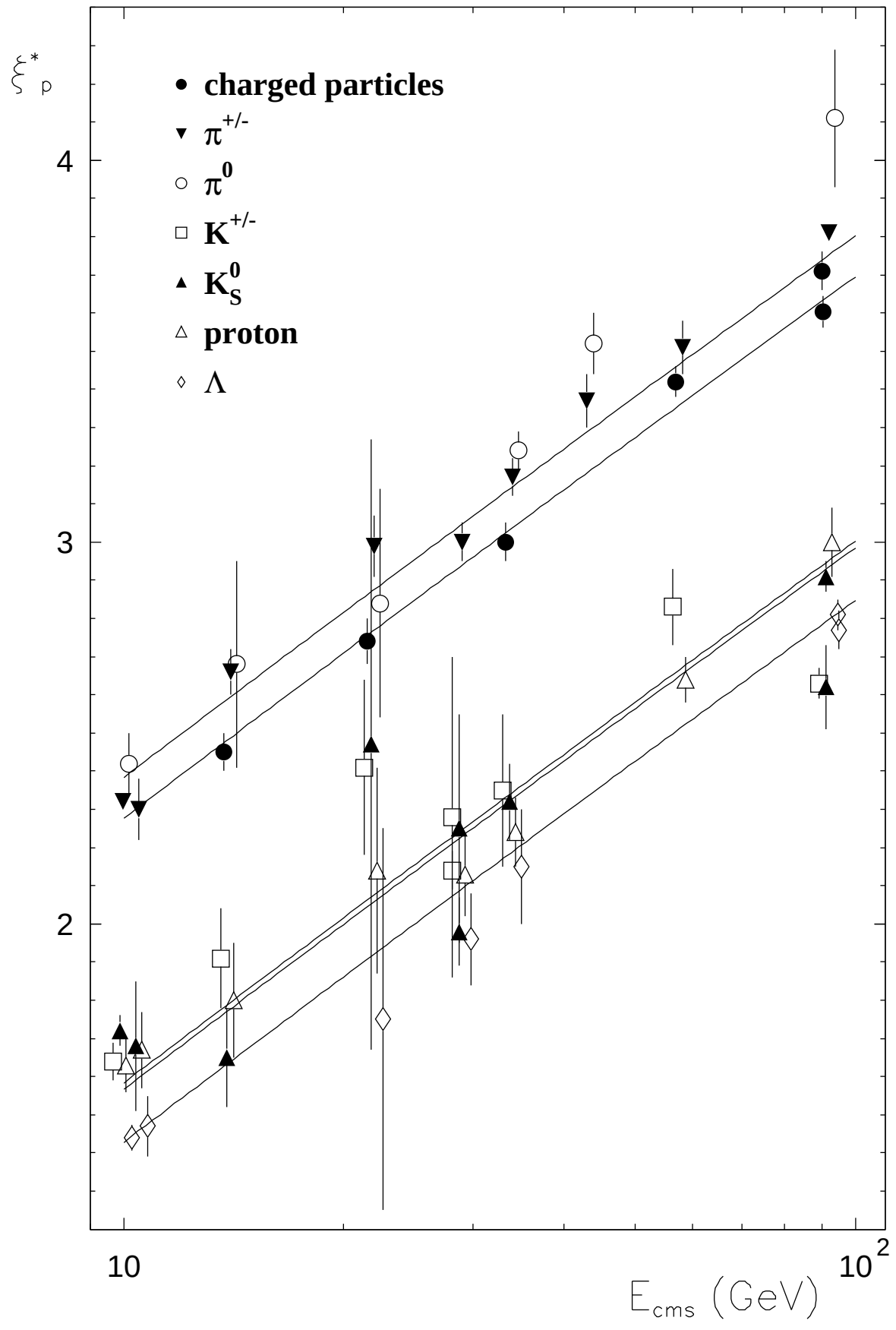

**Figure 4:** *The experimental values of $\xi_p^*$ versus $\sqrt{s}$ for various identified hadrons. The curves are from the combined fit of $\Lambda_{\text{eff}}$ and $F(hadron)$ to the predicted dependence (equations (1) and (8)). From top to bottom, the curves correspond to pions ($\pi^{\pm}$, $\pi^0$), all charged particles, kaons ($K^{\pm}$, $K^0$), protons and $\Lambda$ baryons.*

## 5 Dependence of $\xi_p^*$ on $\sqrt{s}$

A more quantitative comparison is possible for the scaling behaviour of $\xi_p^*$, as parametrised by the QCD scale $\Lambda_{\text{eff}}$. The dependence on $\sqrt{s} = 2E_{\text{beam}}$ of the experimental values of $\xi_p^*$ from table 1 is presented in the figure 4.

The experimental values can be fitted easily to the dependence given by equation (1) with $F = 0$, or to a linear dependence on $Y$. The results for different types of hadrons are given in table 2. Both functions are acceptable and relation (1) gives approximately the correct slope, that should be equal for all hadrons.

Before going on, one should note that in this case the quality of the fits can be considered as an internal check of the estimated errors. The $14$ parameter fit to separate linear functions of $\log E$ for each hadron leads to a total $\chi^2/\text{n.d.f.} = 1.35$. One could conclude that some of the (systematic) errors have been slightly underestimated. It can also be noted that the $K^0$ data give the largest contribution to $\chi^2$.

| particle | $a_h$ | $b_h$ | $\chi^2$/n.d.f. | $Q'_0$ | $\chi^2$/n.d.f. | points |
|---|---|---|---|---|---|---|
| charged | $0.75 \pm 0.11$ | $0.644 \pm 0.029$ | $1.24$ | $0.211 \pm 0.007$ | $1.03$ | 6 |
| $\pi^\pm$ | $0.75 \pm 0.08$ | $0.687 \pm 0.026$ | $1.27$ | $0.181 \pm 0.005$ | $1.61$ | 8 |
| $\pi^0$ | $0.72 \pm 0.21$ | $0.728 \pm 0.053$ | $0.52$ | $0.146 \pm 0.008$ | $0.95$ | 6 |
| $\pi^\pm, \pi^0$ | $0.69 \pm 0.07$ | $0.715 \pm 0.024$ | $1.33$ | $0.175 \pm 0.004$ | $2.15$ | 14 |
| $\mathrm{K}^\pm$ | $0.29 \pm 0.11$ | $0.577 \pm 0.035$ | $2.05$ | $0.589 \pm 0.027$ | $2.24$ | 7 |
| $\mathrm{K}^0$ | $0.46 \pm 0.12$ | $0.529 \pm 0.035$ | $2.69$ | $0.612 \pm 0.023$ | $5.91$ | 9 |
| $\mathrm{K}^\pm, \mathrm{K}^0$ | $0.42 \pm 0.07$ | $0.540 \pm 0.022$ | $2.14$ | $0.603 \pm 0.018$ | $4.02$ | 16 |
| $\mathrm{p}$ | $0.33 \pm 0.15$ | $0.562 \pm 0.046$ | $0.33$ | $0.621 \pm 0.032$ | $1.07$ | 7 |
| $\Lambda$ | $0.03 \pm 0.07$ | $0.610 \pm 0.019$ | $0.34$ | $0.770 \pm 0.025$ | $1.37$ | 7 |
| $\mathrm{p}, \Lambda$ | $0.12 \pm 0.07$ | $0.596 \pm 0.018$ | $2.08$ | $0.725 \pm 0.020$ | $2.08$ | 14 |
| total $\chi^2$/n.d.f. | | | $1.35$ | | $1.94$ | |

**Table 2:** *Result of a fit of the experimental values of $\xi_p^*$ in table 1 to a linear function $\xi_p^* = a_h + b_h \cdot \log(E_{cms}/GeV)$ for various identified hadrons. The scale $Q'_0$ is the result of a second fit to the function $\xi_p^* = \frac{1}{2}Y + \sqrt{CY} - C$, where $Y \equiv \log(E_{beam}/Q'_0)$ and $C = 0.2915$. The fits improve slightly when data at the $\Upsilon$ are excluded. To compare the overall quality of the fits to the two functions, a total $\chi^2$/n.d.f. is given, based on the fits for $\pi^\pm$, $\pi^0$, $\mathrm{K}^\pm$, $\mathrm{K}^0$, $\mathrm{p}$, and $\Lambda$.*

In the fit to equation (1) it is not easy to disentangle $\Lambda_{\text{eff}}$ from $F(Q_0)$, since they are strongly correlated. The simple way out was to constrain $\Lambda_{\text{eff}} = Q_0 = Q'_0$, so that $F(\lambda) = 0$. The fitted values of $Q'_0$ are not equal to the corresponding values of $Q_0$ shown in figure 3. The values of $Q'_0$ are quite near to the particle mass for mesons, but for the baryons $Q'_0$ is significantly smaller.

Figure 5 shows the dependence of $Q'_0$ on (a) the hadron mass and (b) the hadron mass per valence quark. Neither of these plots shows a clear linear dependence. It is possible that this is again due to $F$ in equation (1).

An alternative to setting $F = 0$ is to perform a simultaneous fit of equation (1) to all the data from table 1, where $\Lambda_{\text{eff}}$ is a universal scale, while $F$ is allowed to have different values for hadrons, pions, kaons and the two baryons.

Such a six parameter fit was performed by minimising $\chi^2$ using the MINUIT [24] program, with zero as starting values for $F_h$. There is a strong correlation in the fit of $\log \Lambda_{\text{eff}}$ and $\bar{F}$, the average of $F_h$. Therefore it is practical to fit $\bar{F}$ and differences $F_h - \bar{F}$. The fit has $44$ degrees of freedom and leads to $\chi^2/44 = 1.98$, corresponding to a probability of $1.1 \cdot 10^{-4}$. This value of $\chi^2$ can also be compared to the values of $\chi^2$/n.d.f in table 2, where separate fits for every hadron type were done. This comparison shows that one should probably not reject equation (1) on the basis of the low probability, because it cannot be excluded that the errors of $\xi_p^*$ in table 1 have been underestimated by a factor of $\sqrt{\chi^2/\text{n.d.f.}} = 1.41$, on average. Figure 4 shows that the fitted functions are very near to straight lines and that they follow the data points rather well. Addition of an $\mathcal{O}(Y^{3/2})$ term as in equation (1) does not improve the fit significantly.

The resulting parameters and their external errors are given by:

$$\begin{aligned}
\Lambda_{\text{eff}} &= 0.07\,^{+0.09}_{-0.05}\ \mathrm{G}e\mathrm{V} \\
\bar{F} &= -1.14\,^{+0.53}_{-0.70} \\
F_\pi - \bar{F} &= 0.54 \pm 0.02 \\
F_{\text{charged}} - \bar{F} &= 0.43 \pm 0.02 \\
F_{\mathrm{K}} - \bar{F} &= -0.26 \pm 0.02 \\
F_p - \bar{F} &= -0.28 \pm 0.03 \\
F_\Lambda - \bar{F} &= -0.42 \pm 0.02
\end{aligned} \tag{8}$$

The correlation coefficient of $\Lambda_{\text{eff}}$ and $\bar{F}$ is $100\%$, while the values of $F_h - \bar{F}$ are constrained to have a zero sum.

# 6 The theoretical meaning of $\Lambda_{\text{eff}}$

The determined value of $\Lambda_{\text{eff}}$ from (8) is related to the (running) strong coupling constant by:

$$\frac{\alpha_s}{2\pi} = \frac{1}{b\,Y} = \frac{1}{b\log(E_{\text{beam}}/\Lambda_{\text{eff}})}, \tag{9}$$

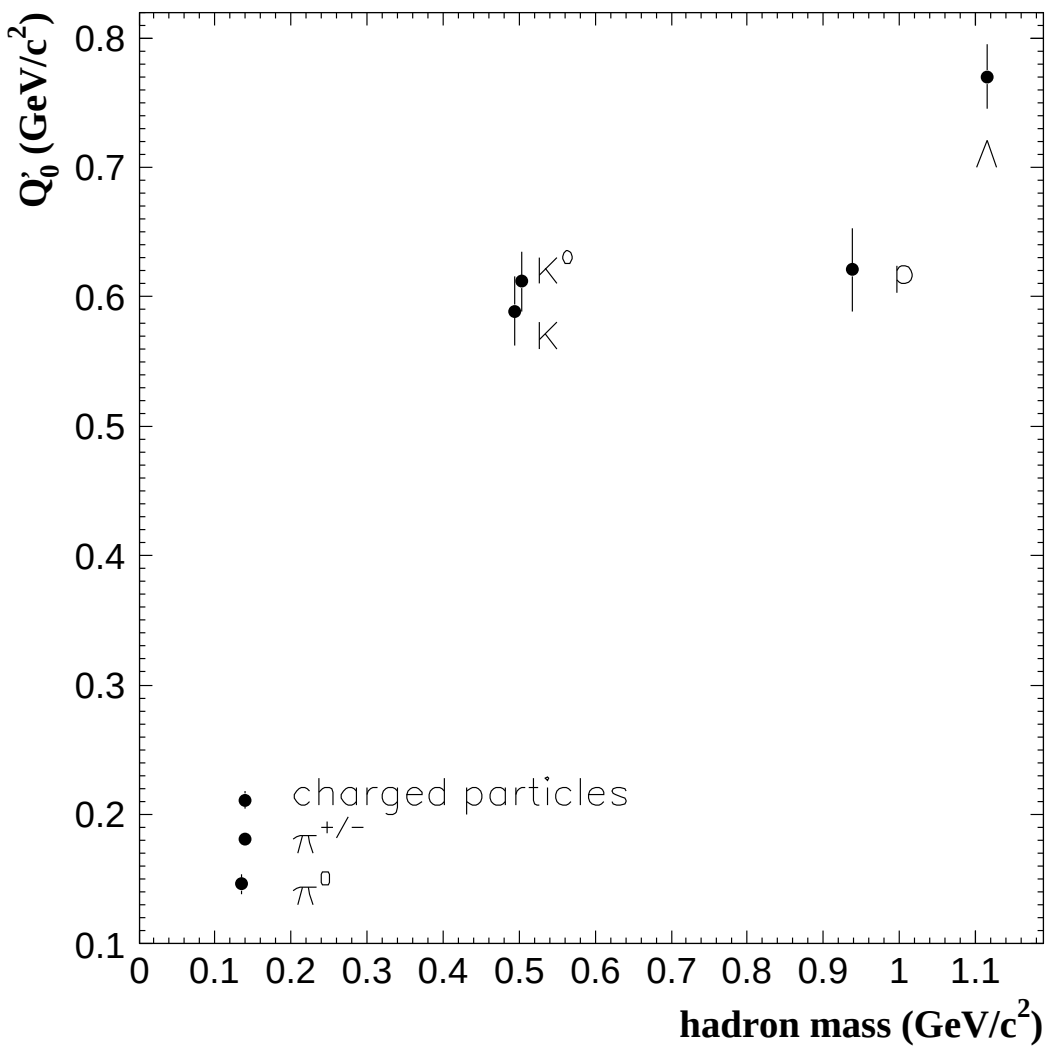

(a) $Q_0'$ versus the hadron mass

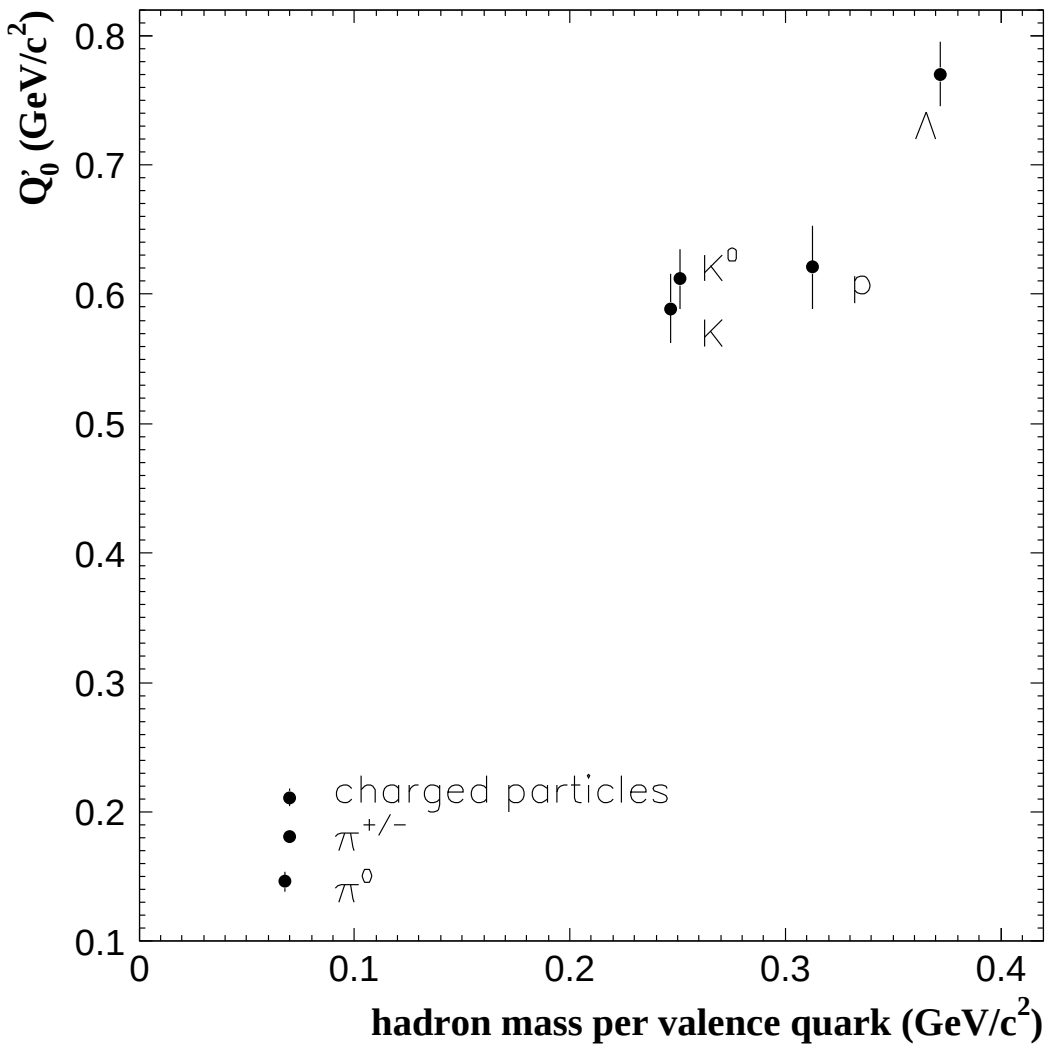

(b) $Q_0'$ versus the hadron mass per quark

**Figure 5:** *The values of $Q_0'$ from table from 2 versus (a) the hadron mass, and (b) the mass per valence quark.*

giving at the $\mathrm{Z}^0$:

$$\alpha_s(\mathrm{LEP}) = 0.108 \pm 0.018 \tag{10}$$

This value is compatible with the more accurate determinations done e.g. at LEP. This would suggest that LPHD + MLLA can really give a quantitative description of the low $x_p$ data, based on perturbative QCD.

However, theoretically the QCD scale $\Lambda_{\text{eff}}$ of MLLA is not well defined. This means that the value of $\Lambda_{\text{eff}}$ as determined in (8) can not be compared to known values of e.g. $\Lambda_{\overline{\text{MS}}}$. A measurement of $\Lambda_{QCD}$ in a well defined subtraction scheme is equivalent to a measurement of $\alpha_s(E_{\text{beam}})$ at a well defined energy scale. To fix this energy scale one would need the $\mathcal{O}(Y^{3/2})$-term in equation 1.

To see this [23], one can express $Y$ in equation (1) in terms of the running coupling $\alpha_s(E_{\text{beam}})$. Using equation (9), the scale dependence of equation (1) translates to:

$$\frac{\mathrm{d}\xi_p^*(E_{\text{beam}}, Q_0)}{\mathrm{d}\log E_{\text{beam}}} = \frac{1}{2} + \frac{a}{8N_c}\sqrt{\frac{N_c\alpha_s}{2\pi}} + \mathcal{O}\left(\alpha_s^{3/2}\right) \tag{11}$$

The $\sqrt{\alpha_s}$ term is the next-to-leading (MLLA) correction to the leading term. The next-to-next-to-leading term $\mathcal{O}(\alpha_s)$ vanishes. An uncertainty of the energy scale would lead to e.g. $\alpha_s(2E)$ instead of $\alpha_s(E)$, inducing a correction in equation (11) of $\mathcal{O}\left(\alpha_s^{3/2}(E)\right)$. This shows that the $\mathcal{O}\left(\alpha_s^{3/2}(E)\right)$ needs to be known for the energy scale to be defined exactly. This is the reason for talking about the scale $\Lambda_{\text{eff}}$, in stead of e.g. $\Lambda_{\overline{\text{MS}}}$.

## 7 Conclusion

The data of LEP are giving a wealth of information about the fragmentation of quarks and gluons, as well as the mechanisms at play in the hadronisation. Although Monte Carlo models like `JETSET` and `HERWIG` can describe these data very accurately, it is questionable whether these models with all their free parameters lead to a better understanding of the physics behind the formation of hadrons in a jet, or only to a better parametrisation.

The 'LPHD + MLLA' approach [1–4] is an interesting attempt to understand some properties of multihadron production in jets in terms of perturbative QCD. This approach tries to take perturbative QCD as near as possible to the limits posed by confinement. It gives predictions for properties of momentum spectra that can be compared with experimental data, but concentrating on observables that can be described without too many unknown free parameters.

The experimental values of the parameter $\xi_p^*$ for various hadrons and at various centre of mass energies have been determined from the published momentum spectra. Subsequently, an investigation was made of the dependence of $\xi_p^*$ on both the centre of mass energy and the mass of the identified hadrons, and this could be compared with the theoretical predictions.

The dependence of $\xi_p^*$ on the centre of mass energy can be described adequately by the MLLA calculations. It is perhaps striking that the low $x_p$ data, that are so intimately related to confinement, can not only be described by perturbative QCD, but can even be used to extract a consistent value of the strong coupling constant. The dependence of $\xi_p^*$ on the mass and flavour of the identified hadron raises some questions that have not yet been answered in the context of LPHD + MLLA and the description of fragmentation by truncated parton cascades.

## Acknowledgements

The author thanks NIKHEF-H for their hospitality while this work was finished and Bert Koene and Paul Kooijman for their critical comments. He is grateful for helpful comments and suggestions from Yuri L. Dokshitzer and Valery A. Khoze.